\documentclass[12pt,showkeys,showpacs]{article}

\usepackage[T2A]{fontenc}
\usepackage[english]{babel}
\usepackage{amsmath}
\usepackage{amssymb}
\newcommand{\N}{N\raise.7ex\hbox{\underline{$\circ $}}$\;$}
\begin{document}

\title{\textbf{Generalized boosts with shell structure \\ of the parameter space}}
\author{\textbf{A.~N.~Tarakanov}
\thanks{E-mail: tarak-ph@mail.ru }\\
{\small \textit{Institute of Informational Technologies,}} \\
{\small \textit{Belarusian State University of Informatics and Radioelectronics}} \\
{\small \textit{Kozlov str. 28, 220037, Minsk, Belarus}} }

\date{}
\maketitle

\begin{abstract}
A modification of boost transformation in arbitrary pseudo-Euclidean space is suggested, which in the case of the Minkowsky space admits an existing of inertial reference frames moving with velocities taking values in a certain bounded interval. The velocity space may be partitioned by hypersurfaces $\boldsymbol{\beta}^2 = \beta^2_{\mathrm{k}} = \mathrm{const}$, $\mathrm{k} = 1,2,3,...$, into a finite or countable number of domains (shells), each of which has own class of inertial `reference frames' and the velocity composition law. These shells are in one-to-one correspondence. A set of mappings of shells to each other forms the group, isomorphic to permutation group $\mathbb{Z}_N$ in the case of finite ($N+1$) number of shells, or the group of integers $\mathbb{Z}$ in the case of countable number of shells in the velocity space.
\end{abstract}


\vspace {10mm}

\textbf{1.} As is well known, the Lorentz transformations can be derived in a purely mathematical way without involving any physical ideas. The metaphysical assumption that space and time constitute a single space-time continuum is in fact a mathematical definition of the Minkowski space, $\mathbf{E}^{\mathbb{R}}_{1,3}$, in which the infinitesimal distance between two points (events) is determined by an interval
\begin{equation}
ds^2 = \eta_{\mu\nu} dx^{\mu} dx^{\nu} \, , \qquad \boldsymbol{\eta} = \{ \eta_{\mu\nu} \} = \mathrm{diag}(1,-1,-1,-1) \, , \label{1}
\end{equation}
so that all properties of space-time are determined by the metric tensor $\eta_{\mu\nu}$.

The Lorentz transformations in $\mathbf{E}^{\mathbb{R}}_{1,3}$ give a correspondence between quantities (such as vectors and tensors) considered from the point of view of different inertial reference frames (i. r. f.), which relative velocity is variable in the limits either $0 \leq \boldsymbol{\beta}^2 = \mathbf{v}^2 / c^2 < 1$ (subluminal transformations), or $1 < \boldsymbol{\beta}^2 < \infty$ (superluminal transformations). It had been repeatedly pointed out that superluminal Lorentz transformations can be obtained from conventional subluminal ones by conformal transformation of the velocity that constitutes so-called \textit{the principle of duality} (see, e. g., review \cite{Rec}). We may consider a structure of the velocity space firmly established in the subluminal domain. It is three-dimensional space of constant curvature realized inside the sphere $\boldsymbol{\beta}^2 = 1$. But as to its structure beyond the light barrier, it is impossible to make so certain conclusions now. It means that standard superluminal modification of Special Relativity is not unique. Particularly, it is possible to raise the question of search of such pseudo-orthogonal transformations which would admit variation of the velocity from $c$ to some finite value $C>c$. It is shown in \cite{Tar1} that Lorentz boosts are naturally generalized to ($p+q$)-dimensional pseudo-Euclidean space, denoted further by $\mathbf{E}^{\mathbb{R}}_{p,q}$ in accordance with \cite{Tar2}. Below we suggest a generalization of Lorentz boosts in $\mathbf{E}^{\mathbb{R}}_{p,q}$, which parameters have both upper and lower limits for the velocity.

\textbf{2.} Let's briefly remind the main points of the paper \cite{Tar1}. Let $x^{\mu}$, $\mu = 1,2,...,p+q$, be coordinates, covering $\mathbf{E}^{\mathbb{R}}_{p,q}$, and $x^{\mathrm{m}}$ ($m$ fixed) has the meaning of time. Then real boosts, $x'^{\mu} = L^{\mu}_{\, \cdot \, \nu} x^{\nu}$, are given by
\begin{equation}
\mathbf{\hat{L}} = \{L^{\mu}_{\, \cdot \, \nu} \} = \mathbf{\hat{1}}_{p+q} + \gamma \frac{\mathbf{\hat{v}}}{c} + \frac{\gamma-1}{\boldsymbol{\beta}^2_v} \frac{\mathbf{\hat{v}}^2}{c^2} \, , \label{2}
\end{equation}
where
\begin{equation}
\gamma = (1 - \boldsymbol{\beta}^2_v)^{-1/2} \, , \qquad 0 \leq \boldsymbol{\beta}^2_v = -\frac{\eta^{\mathrm{mm}}}{c^2} \eta_{AB} v^A v^B < 1 \, ; \label{3}
\end{equation}
\begin{equation}
\boldsymbol{\eta} = \{ \eta_{\mu\nu} \} = \mathrm{diag}(\underbrace{+1,...,+1}_p,\underbrace{-1,...,-1}_q) \, , \;\; A,B = 1,2,...,m-1,m+1,...,p+q \, ; \label{4}
\end{equation}
\begin{equation}
(\mathbf{\hat{v}})^{\mu}_{\, \cdot \, \nu} = v_A (\eta^{A\mu}\delta^{\mathrm{m}}_\nu - \eta^{\mathrm{m}\mu}\delta^A{\nu}) \, , \qquad v_A = \eta_{AB} v^B \, . \label{5}
\end{equation}

In the Minkowsli space, $\mathbf{E}^{\mathbb{R}}_{1,3}$, $v^A \equiv v^i$ ($A=i=1,2,3$) are components of the relative velocity of two i. r. f. and eq.~\eqref{2} is well-known Lorentz boosts in  arbitrary direction. Transformation \eqref{2} in $\mathbf{E}^{\mathbb{R}}_{p,q}$ has a meaning also for non-positive values of $\boldsymbol{\beta}^2_v$: $-\infty < \boldsymbol{\beta}^2_v < 0$ and $\boldsymbol{\beta}^2_v=0$, and in the latter case eq. \eqref{2} degenerates into
\begin{equation}
\mathbf{\hat{L}} = \mathbf{\hat{1}}_{p+q} + \frac{\mathbf{\hat{v}}}{c} + \frac{\mathbf{\hat{v}}^2}{2c^2}  \, . \label{6}
\end{equation}

The application to eq. \eqref{2} of a conformal transformation in the velocity space
\begin{equation}
v^A = \frac{u^A}{\boldsymbol{\beta}^2_u} \; , \label{7}
\end{equation}
leads to transformation
\begin{equation}
\mathbf{\hat{L}} = \mathbf{\hat{1}}_{p+q} + \frac{\Gamma}{\boldsymbol{\beta}^2_u} \frac{\mathbf{\hat{u}}}{c} + \frac{\Gamma - 1}{\boldsymbol{\beta}^2_u} \frac{\mathbf{\hat{u}}^2}{c^2} \, , \label{8}
\end{equation}
which is generalization of superluminal boosts of the Minkowski space $\mathbf{E}^{\mathbb{R}}_{1,3}$ to the space $\mathbf{E}^{\mathbb{R}}_{p,q}$. Here
\begin{equation}
\Gamma = (1 - \boldsymbol{\beta}^{-2}_u)^{-1/2} \, , \qquad 1 < \boldsymbol{\beta}^2_u = -\frac{\eta^{\mathrm{mm}}}{c^2} \eta_{AB} u^A u^B < \infty \, . \label{9}
\end{equation}

Transformation \eqref{7} maps the domain $\boldsymbol{\beta}^2_v \leq 0$ into itself, transforms ($p+q-1$)-dimensional cone $\boldsymbol{\beta}^2_v = 0$ to infinity $\boldsymbol{\beta}^2_u = \infty$, and the domain $0 \leq \boldsymbol{\beta}^2_v < 1$ into domain $1 < \boldsymbol{\beta}^2_u < \infty$. Thus, union of domains $\boldsymbol{\beta}^2_v \leq 0$, $\boldsymbol{\beta}^2_v = 0$, $0 \leq \boldsymbol{\beta}^2_v < 1$, $\boldsymbol{\beta}^2_v = 1$ and $1 < \boldsymbol{\beta}^2_u < \infty$ is the whole velocity space, which will hereinafter be denoted by $\mathbf{V}^{\mathbb{R}}_{p,q}$, said domains by $\mathbf{V}^{-}_{p,q}$, $\mathbf{V}^{0}_{p,q}$, $\mathbf{V}^{<}_{p,q}$, $\mathbf{V}^1_{p,q}$ and $\mathbf{V}^{>}_{p,q}$, respectively, and $\mathbf{V}^{+}_{p,q} = \mathbf{V}^{<}_{p,q} \bigcup \mathbf{V}^1_{p,q} \bigcup \mathbf{V}^{>}_{p,q}$ be a space, where $\boldsymbol{\beta}^2_v \geq 0$. In $\mathbf{E}^{\mathbb{R}}_{1,3}$ there remain only domains $\mathbf{V}^{<}_{1,3}$, $\mathbf{V}^{>}_{1,3}$, which are the arenas of Special Relativity and its superluminal generalization, respectively, and $\mathbf{V}^1_{1,3}$.

If we go to ($p+q-1$)-dimensional space $\mathbf{V}^{\mathbb{R}}_{p,q}$, then transformations \eqref{2} and \eqref{8} are defined inside and outside the hypersurface $\boldsymbol{\beta}^2 = 1$, which is a pseudo-Riemannian hyperbolic space of constant curvature $H_{p,q}$. Transformation \eqref{7} conserves this hypersurface invariant and maps its interiority to its exteriority and vice versa (see~\cite{Tar2}). Hypersurface $\boldsymbol{\beta}^2 = 1$ divides the space $\mathbf{V}^{+}_{p,q}$  into two parts $\mathbf{V}^{<}_{p,q} = D^H_{p,q}$ and $\mathbf{V}^{>}_{p,q} = \tilde{D}^H_{p,q}$, and eqs. \eqref{2}, \eqref{8} on it degenerate into $\mathbf{\hat{L}} = \mathbf{\hat{1}}_{p+q} - \mathbf{\hat{v}}^2/c^2$. Thus, the modification of transformations \eqref{2}, \eqref{8} must consist in mapping of $\mathbf{V}^{<}_{p,q}$ into a space bounded by two similar hypersurfaces of singularity:
\begin{equation}
\boldsymbol{\beta}^2 = \beta^2_1 = C^2_1 / c^2 \qquad \hbox{and} \qquad \boldsymbol{\beta}^2 = \beta^2_2 = C^2_2 / c^2 \, , \label{10}
\end{equation}
and besides the ratio between $C_1$ and $c$ can be arbitrary. We call such a domain the shell. The required mapping can be written in the following form
\begin{equation}
v^A = \frac{u^A}{\boldsymbol{\beta}^2_u \sqrt{\beta^2_2 - \beta^2_1}} \left[ \sqrt{\boldsymbol{\beta}^2_u (\boldsymbol{\beta}^2_u - 1)} - \beta_1 \sqrt{\beta^2_2 - \boldsymbol{\beta}^2_u} \right] \, , \label{11}
\end{equation}
with
\begin{equation}
\beta^2_1 < \boldsymbol{\beta}^2_u < \beta^2_2 \, , \qquad \hbox{or} \qquad C^2_1 < \mathbf{u}^2 < C^2_2 \, . \label{12}
\end{equation}

We find from eq.~\eqref{12}
\begin{equation}
\boldsymbol{\beta}^2_v = \frac{1}{\boldsymbol{\beta}^2_u (\beta^2_2 - \beta^2_1)} \left[ \boldsymbol{\beta}^4_u - 2\beta^2_1 \boldsymbol{\beta}^2_u + \beta^2_1 \beta^2_2 - 2\beta_1 \sqrt{\boldsymbol{\beta}^2_u (\boldsymbol{\beta}^2_u - \beta^2_1)(\beta^2_2 - \boldsymbol{\beta}^2_u)} \right] \, , \label{13}
\end{equation}
\begin{equation}
\gamma = (1 - \boldsymbol{\beta}^2_v)^{-1/2} = \sqrt{\frac{\boldsymbol{\beta}^2_u (\beta^2_2 - \beta^2_1)}{(\boldsymbol{\beta}^2_u - \beta^2_1)(\beta^2_2 - \boldsymbol{\beta}^2_u) - 4\beta^2_1 \boldsymbol{\beta}^2_u} \left[ 1- \sqrt{\frac{4\beta^2_1 \boldsymbol{\beta}^2_u}{(\boldsymbol{\beta}^2_u - \beta^2_1)(\beta^2_2 - \boldsymbol{\beta}^2_u)}} \right]} \, . \label{14}
\end{equation}
Substituting eqs.~\eqref{11}, \eqref{13} and \eqref{14} into eq.~\eqref{2} yields the final expression for the boost transform acting inside the shell \eqref{12} under consideration:
\begin{equation}
\mathbf{\hat{L}} = \mathbf{\hat{1}}_{p+q} + \frac{\gamma \left(\sqrt{\boldsymbol{\beta}^2_u(\boldsymbol{\beta}^2_u - \beta^2_1)} - \beta_1 \sqrt{\beta^2_2 - \boldsymbol{\beta}^2_u} \right)}{\boldsymbol{\beta}^2_u \sqrt{\beta^2_2 - \beta^2_1}} \frac{\mathbf{\hat{u}}}{c} + \frac{\gamma - 1}{\boldsymbol{\beta}^2_u} \frac{\mathbf{\hat{u}}^2}{c^2} \, . \label{15}
\end{equation}

It is easy to see that in the limit $C_1 \rightarrow 0$, $C_2 \rightarrow \infty$, (i.e. $\beta_1 \rightarrow 0$, $\beta_2 \rightarrow 1$), the Lorentz-Fitzgerald factor \eqref{14} tends to $\gamma = (1 - \boldsymbol{\beta}^2_u)^{-1/2}$, transformation \eqref{11} becomes the identity mapping, $v^A=u^A$, and eq. \eqref{15} reduces to eq. \eqref{2}, which corresponds to subluminal boosts. In the other limit, $C_1 \rightarrow c$, $C_2 \rightarrow \infty$ (i.e. $\beta_1 \rightarrow 1$, $\beta_2 \rightarrow \infty$), the factor \eqref{14} tends to $\Gamma = (1 - \boldsymbol{\beta}^{-2}_u)^{-1/2}$, eq. \eqref{11} degenerates into a conformal transformation \eqref{7} in the opposite direction, i.e.  $v^A = -u^A / \boldsymbol{\beta}^2_u$, whereas eq. \eqref{15} reduces to the expression \eqref{8}, corresponding to superluminal boosts in the opposite direction. Thus, because of condition \eqref{12} expression \eqref{15} in the case of the Minkowski space $\mathbf{E}^\mathbb{R}_{1,3}$ describes transformations of i. r. f., that can move with velocities greater than $C_1$ and smaller than $C_2$.

Eq. \eqref{15} leads to a composition of velocities law
\begin{equation}
\begin{split}
V'^A &= c \frac{L^A_{\, \cdot \, B} V^B + c L^A_{\, \cdot \, m}}{L^m_{\, \cdot \, B} V^B + c L^m_{\, \cdot \, m}} = {} \\ &= \frac{V^A - (\gamma - 1)\frac{(\mathbf{u} \cdot \mathbf{V})}{c^2 \boldsymbol{\beta}^2_u} u^A + \frac{\gamma \left(\sqrt{\boldsymbol{\beta}^2_u(\boldsymbol{\beta}^2_u - \beta^2_1)} - \beta_1 \sqrt{\beta^2_2 - \boldsymbol{\beta}^2_u} \right)}{\boldsymbol{\beta}^2_u \sqrt{\beta^2_2 - \beta^2_1}} u^A}{\gamma \left[1 + \frac{\sqrt{\boldsymbol{\beta}^2_u(\boldsymbol{\beta}^2_u - \beta^2_1)} - \beta_1 \sqrt{\beta^2_2 - \boldsymbol{\beta}^2_u}}{\sqrt{\beta^2_2 - \beta^2_1}}\frac{(\mathbf{u} \cdot \mathbf{V})}{c^2 \boldsymbol{\beta}^2_u} \right]}\, , \label{16}
\end{split}
\end{equation}
where $(\mathbf{u} \cdot \mathbf{V}) = -\eta^{\mathrm{mm}}\eta_{AB} u^A V^B$ is a scalar product of two ($p+q-1$)-dimensional vectors, $V^A = c dx^A/dx^m$ and $V'^A = c dx'^A/dx'^m$ are velocities of the material point in i.~r.~f.~$\mathrm{K}$ and $\mathrm{K}'$, moving with respect to each other with velocity $u^A$. In the limit $\beta_1 \rightarrow 0$, $\beta_2 \rightarrow 1$ the law \eqref{16} reduces to
\begin{equation}
V'^A = \frac{V^A + u^A + (\gamma - 1) \left( 1- \frac{(\mathbf{u} \cdot \mathbf{V})}{c^2 \boldsymbol{\beta}^2_u} \right) u^A}{\gamma \left(1 + \frac{(\mathbf{u} \cdot \mathbf{V})}{c^2} \right)}\, , \label{17}
\end{equation}
and when $\beta_1 \rightarrow 1$, $\beta_2 \rightarrow \infty$ it looks like
\begin{equation}
V'^A = \frac{V^A + (1/\Gamma -1) u^A - (\Gamma - 1) \left( 1 + \frac{(\mathbf{u} \cdot \mathbf{V})}{c^2 \boldsymbol{\beta}^2_u} \right) u^A}{(\Gamma - 1/\Gamma) \left(1 - \frac{(\mathbf{u} \cdot \mathbf{V})}{c^2} \right) + 1/\Gamma}\, . \label{18}
\end{equation}

\textbf{3.} Transformations described above can be generalized to the case when the velocity space $\mathbf{V}^\mathbb{R}_{p,q}$ contains a finite ($N$) or countable number of hypersurfaces $\boldsymbol{\beta}^2 = \beta^2_\mathrm{k}$ that divide it into a finite ($N+1$) or countable number of shells, and in the k-th shell $\boldsymbol{\beta}^2$ varies within the limits $\beta^2_{\mathrm{k}-1} \leq \boldsymbol{\beta}^2 < \beta^2_\mathrm{k}$, where the equality sign holds only for $\mathrm{k}=1$, $\beta_0 = 0$. For finite number of shells $\mathrm{k}=1,2,...,\mathrm{N}$ and $\beta^2_\mathrm{N} = \infty$, and $\mathrm{k}=1,2,...$, for countable number. Then each shell admits action of pseudo-orthogonal transformations consisting of ($p+q$)-dimensional rotations (and pseudo-rotations) and ($p+q$)-dimensional boosts of the type \eqref{15}:
\begin{equation}
\mathbf{\hat{L}}_{\mathrm{k}} = \mathbf{\hat{1}}_{p+q} + \frac{\gamma_{\mathrm{k}} \left(\sqrt{\boldsymbol{\beta}^2_u(\boldsymbol{\beta}^2_u - \beta^2_{\mathrm{k}-1})} - \beta_{\mathrm{k}-1} \sqrt{\beta^2_{\mathrm{k}} - \boldsymbol{\beta}^2_u} \right)}{\boldsymbol{\beta}^2_u \sqrt{\beta^2_{\mathrm{k}} - \beta^2_{\mathrm{k}-1}}} \frac{\mathbf{\hat{u}}}{c} + \frac{\gamma_{\mathrm{k}} - 1}{\boldsymbol{\beta}^2_u} \frac{\mathbf{\hat{u}}^2}{c^2} \, , \label{19}
\end{equation}
where $\beta_{\mathrm{k}} = C_{\mathrm{k}}/c$,
\begin{equation}
\begin{split}
\gamma_{\mathrm{k}} &= (1 - \boldsymbol{\beta}^2_u)^{-1/2} = {} \\ &= \sqrt{\frac{\boldsymbol{\beta}^2_u (\beta^2_{\mathrm{k}} - \beta^2_{\mathrm{k}-1})}{(\boldsymbol{\beta}^2_u - \beta^2_{\mathrm{k}-1})(\beta^2_{\mathrm{k}} - \boldsymbol{\beta}^2_u) - 4\beta^2_{\mathrm{k}-1} \boldsymbol{\beta}^2_u} \left[ 1- \sqrt{\frac{4\beta^2_{\mathrm{k}-1} \boldsymbol{\beta}^2_u}{(\boldsymbol{\beta}^2_u - \beta^2_{\mathrm{k}-1})(\beta^2_{\mathrm{k}} - \boldsymbol{\beta}^2_u)}} \right]} \, . \label{20}
\end{split}
\end{equation}

Between the shells under consideration it is possible to establish a one-to-one correspondence, which, therefore, will generalize the above-mentioned principle of duality. If the maps of the first domain into the k-th and m-th shells are defined by expressions
\begin{equation}
v^A = \frac{u^A}{\boldsymbol{\beta}^2_u \sqrt{\beta^2_{\mathrm{k}} - \beta^2_{\mathrm{k}-1}}} \left[ \sqrt{\boldsymbol{\beta}^2_u (\boldsymbol{\beta}^2_u - \beta^2_{\mathrm{k}-1})} - \sqrt{\beta^2_{\mathrm{k}-1} (\beta^2_{\mathrm{k}} - \boldsymbol{\beta}^2_u} \right] \, , \label{21}
\end{equation}
\begin{equation}
v^A = \frac{w^A}{\boldsymbol{\beta}^2_w \sqrt{\beta^2_{\mathrm{m}} - \beta^2_{\mathrm{m}-1}}} \left[ \sqrt{\boldsymbol{\beta}^2_w (\boldsymbol{\beta}^2_w - \beta^2_{\mathrm{m}-1})} - \sqrt{\beta^2_{\mathrm{m}-1} (\beta^2_{\mathrm{m}} - \boldsymbol{\beta}^2_w} \right] \, , \label{22}
\end{equation}
respectively, where $\boldsymbol{\beta}^2_w = \mathbf{w}^2/c^2$, then the map from the k-th shell into m-th one is given by
\begin{equation}
u^A = \sqrt{\frac{\boldsymbol{\beta}^2_u}{\boldsymbol{\beta}^2_w}} w^A \, , \label{23}
\end{equation}
where $\beta_u$ being function of $\beta_w$, can be determined from an algebraic equation of fourth degree.

If we denote
\begin{equation}
\mathrm{X} = \frac{\boldsymbol{\beta}^2_u}{\beta_{\mathrm{k}-1} \beta_{\mathrm{k}}} \, , \qquad D = \frac{\beta_{\mathrm{k}-1}}{\beta_{\mathrm{k}}} = \frac{C_{\mathrm{k}-1}}{C_{\mathrm{k}}} \, , \label{24}
\end{equation}
\begin{equation}
B = (D^{-1} - D) \boldsymbol{\beta}^2_v = \frac{D^{-1} - D}{\boldsymbol{\beta}^2_w (\beta_{\mathrm{m}} - \beta_{\mathrm{m}-1})} \left[ \sqrt{\boldsymbol{\beta}^2_w (\boldsymbol{\beta}^2_w - \beta^2_{\mathrm{m}-1})} - \sqrt{\beta^2_{\mathrm{m}-1} (\beta^2_{\mathrm{m}} - \boldsymbol{\beta}^2_w} \right]^2 \, , \label{25}
\end{equation}
then equation for $\mathrm{X}$ looks like
\begin{equation}
\mathrm{X}^4 - 2B\mathrm{X}^3 + (B^2 +4BD -2)\mathrm{X}^2 - 2B\mathrm{X} +1 = 0 \, , \label{26}
\end{equation}
and its solutions are
\begin{equation}
\mathrm{X}_{1,3} = \frac{B \pm 2 \sqrt{1 - BD}}{2} \left[ 1+ \sqrt{1 - \frac{4}{[B \pm 2 \sqrt{1 - BD}]^2}} \right]\, , \label{27}
\end{equation}
\begin{equation}
\mathrm{X}_{2,4} = \frac{1}{\mathrm{X}_{1,3}} \, . \label{28}
\end{equation}

From these solutions one should choose only those that satisfy the condition \eqref{12}, or $D < \mathrm{X} < D^{-1}$. It follows from eqs. \eqref{3} and \eqref{25} that $D^2 \leq 1 - BD < 1$, and the condition of reality of the solution leads to inequalities
\begin{equation}
\frac{B \pm 2 \sqrt{1 - BD}}{2} > 1 \, ,  \qquad \hbox{if} \qquad \mathrm{X} > 0 \, , \boldsymbol{\beta}^2_u > 0 \, , \label{29}
\end{equation}
or
\begin{equation}
\frac{B \pm 2 \sqrt{1 - BD}}{2} < -1 \, ,  \qquad \hbox{if} \qquad \mathrm{X} < 0 \, , \boldsymbol{\beta}^2_u > 0 \, . \label{30}
\end{equation}

Obviously, condition \eqref{30}, corresponding to solutions $\mathrm{X}_3$ and $\mathrm{X}_4$, is unacceptable, since $\mathrm{X}>0$ due to condition \eqref{12}. Thus, \eqref{23} gives two solutions
\begin{equation}
u^A = \sqrt{\frac{\beta_{\mathrm{k}-1} \beta_{\mathrm{k}} \mathrm{X}_1 (\boldsymbol{\beta}^2_w)}{\boldsymbol{\beta}^2_w}} w^A \label{31}
\end{equation}
and
\begin{equation}
u^A = \sqrt{\frac{\beta_{\mathrm{k}-1} \beta_{\mathrm{k}}}{\boldsymbol{\beta}^2_w \mathrm{X}_1 (\boldsymbol{\beta}^2_w)}} w^A \, , \label{32}
\end{equation}
which are inverse to each other in the sense that if \eqref{31} maps the internal boundary $\boldsymbol{\beta}^2 = \beta_{\mathrm{k}-1}$ onto the internal one, $\boldsymbol{\beta}^2 = \beta_{\mathrm{n}-1}$, and the external boundary $\boldsymbol{\beta}^2 = \beta_{\mathrm{k}}$ onto the external one, $\boldsymbol{\beta}^2 = \beta_{\mathrm{n}}$ , then \eqref{32}, on the contrary, maps the internal boundary $\boldsymbol{\beta}^2 = \beta_{\mathrm{k}-1}$ onto the external one, $\boldsymbol{\beta}^2 = \beta_{\mathrm{n}}$, and the external boundary $\boldsymbol{\beta}^2 = \beta_{\mathrm{k}}$ onto the internal one, $\boldsymbol{\beta}^2 = \beta_{\mathrm{n}-1}$. The mappings \eqref{31}-\eqref{32} include the mappings \eqref{7}, \eqref{11}, \eqref{21}, \eqref{22} as the limiting cases at $\beta_{\mathrm{k}-1} \rightarrow 0$, $\beta_{\mathrm{k}} \rightarrow 1$.

\textbf{4.} Now, if we denote the coordinates in k-th shell by $v^A_{\mathrm{k}}$, then the mapping \eqref{31}, \eqref{32} can be written as
\begin{equation}
v^A_{\mathrm{k}} = Z_{\mathrm{kn}}(\boldsymbol{\beta}^2_{v_{\mathrm{n}}}) v^A_{\mathrm{n}} \, . \label{33}
\end{equation}

The transformation of coordinates inside k-th shell is given by the expression \eqref{19}, where $v^A_{\mathrm{k}}$ should be substituted instead of instead of $u^A$, and besides it induces similar transformations in all shells. Indeed,
\begin{equation}
\begin{split}
v'^A_{\mathrm{k}} &= L^A_{\mathrm{k} \cdot B}(v^2_{\mathrm{k}}) v^B_{\mathrm{k}} = L^A_{\mathrm{k} \cdot B}(v^2_{\mathrm{k}}) Z_{\mathrm{kn}}(\boldsymbol{\beta}^2_{v_{\mathrm{n}}}) v^B_{\mathrm{n}} = {} \\ &= Z_{\mathrm{kn}}(\boldsymbol{\beta}^2_{v_{\mathrm{n}}}) L^A_{\mathrm{k} \cdot B}(Z^2_{\mathrm{kn}}(\boldsymbol{\beta}^2_{v_{\mathrm{n}}}) v^2_{\mathrm{n}}) v^B_{\mathrm{n}} = Z_{\mathrm{kn}}(\boldsymbol{\beta}^2_{v_{\mathrm{n}}}) L^A_{\mathrm{n} \cdot B}(v^2_{\mathrm{n}}) v^B_{\mathrm{n}} = Z_{\mathrm{kn}}(\boldsymbol{\beta}^2_{v_{\mathrm{n}}}) v'^A_{\mathrm{n}} \, . \label{34}
\end{split}
\end{equation}

Suppose now that there are two mappings $\mathrm{k} \rightarrow \mathrm{n}$ and $\mathrm{n} \rightarrow \mathrm{m}$. Then we can define the product of these mappings by formula
\begin{equation}
Z_{\mathrm{km}}(\boldsymbol{\beta}^2_{v_{\mathrm{m}}}) = Z_{\mathrm{kn}}(Z^2_{\mathrm{nm}}(\boldsymbol{\beta}^2_{v_{\mathrm{m}}}) \boldsymbol{\beta}^2_{v_{\mathrm{m}}}) v^2_{\mathrm{n}}) Z_{\mathrm{nm}}(\boldsymbol{\beta}^2_{v_{\mathrm{m}}}) \, . \label{35}
\end{equation}
Obviously, there exists the identity mapping $Z_{\mathrm{kk}}(\boldsymbol{\beta}^2_{v_{\mathrm{k}}})=1$, as well as the inverse mapping $Z^{-1}_{\mathrm{kn}} = Z_{\mathrm{nk}}(\boldsymbol{\beta}^2_{v_{\mathrm{k}}})$. Associativity of the multiplication law is evident from eq. \eqref{35}. Therefore, the set of mappings \eqref{33} forms a cyclic group $\mathbb{Z}_N$ of order $N$ in the case of a finite, $N+1$, number of shells, or a group isomorphic to the group $\mathbb{Z}$ of integers if the number of shells is countable.

The mappings \eqref{31}, \eqref{32} can be extended to the domain $\mathbf{V}^{-}_{p,q}$, so that the latter can also be divided into finite or countable number of shells that are characterized by boundaries with  negative $\boldsymbol{\beta}^2_{v_{\mathrm{k}}}$. In this case, condition \eqref{30} will be satisfied, and the mapping from the k-th shell to m-th one will be given by formulae \eqref{31}, \eqref{32}, where $\mathrm{X}_3<0$ should be substituted instead of $\mathrm{X}_1>0$.

In conclusion, we note that the velocity space $\mathbf{V}^{\mathbb{R}}_{p,q}$ is a subspace of the parameter space of the group $O(p,q)$. From this point of view, the mapping \eqref{7}, \eqref{11}, \eqref{21}, \eqref{22}, \eqref{31}, \eqref{32} are reparametrizations of boosts of the group $O(p,q)$, defining the structure of the parameter space. Obviously, for $\mathbf{E}^{\mathbb{R}}_{1,3}$ the space $\mathbf{V}^{\mathbb{R}}_{1,3}$, realized inside the sphere $\boldsymbol{\beta}^2=1$, acquires the physical meaning. The question of the physical meaning of the shell structure of $\mathbf{V}^{\mathbb{R}}_{1,3}$ remains open. We can note only some similarities with the splitting of the front of an electromagnetic wave into Fresnel zones. From infinite set of parametrizations of group $O(p,q)$ (in particular, of group $O(1,3)$), which probably to think up, apparently, some of them can be applied to physical problems.


\begin{thebibliography}{99}

\bibitem{Rec} Recami E. Relativity Theory and its Generalization. // In: Astrofisica e Cosmologia, Gravitazione, Quanti e Relativit\`{a} negli sviluppi del pensiero scientifico di Albert Einstein. -- Firenze: Giunti Barb\`{e}ra, 1979.
\bibitem{Tar1} Tarakanov A. N. Real and Complex ``Boosts'' in Arbitrary Pseudo Euclidean Spaces. // Theor. \& Math. Phys. (USA), 1976, \textbf{18}, no. 3, 838-842.
\bibitem{Tar2} Tarakanov A. N. On Certain Isometric Representations for Multidimensional Spaces of Constant Curvature. // Rep. Math. Phys., 1991, \textbf{29}, no. 2, 195-212.

\end{thebibliography}
\end{document}